\documentclass[aps,pre,superscriptaddress,groupedaddress]{revtex4}  
\usepackage{latexsym}
\usepackage{natbib}
\usepackage{amsmath}
\usepackage{amsfonts}
\usepackage{amssymb}
\usepackage[T1]{fontenc}
\usepackage{color,graphicx}   
\usepackage{verbatim}   
\usepackage{t1enc}
\usepackage{anysize}
\usepackage{subfig}
\usepackage[percent]{overpic} 
\usepackage{float}  
\usepackage{appendix} 
\usepackage{booktabs} 
\usepackage{natbib}
\usepackage{multirow}
\usepackage{array}
\usepackage{color}

\DeclareGraphicsExtensions{.pdf,.pgn,.jpg}
\usepackage{placeins}

\begin{document}


\title{Competing for Attention in Social Media under Information
  Overload Conditions}



\author{Ling Feng}
\affiliation{Department of Physics and Centre for Computational
  Science and Engineering, National University of Singapore, 
  117542, Singapore}

\author{Yanqing Hu}
\affiliation{School of Mathematics, Southwest  Jiaotong University, Chengdu 610031, China}

\author{Baowen Li}
\affiliation{Department of Physics and Centre for Computational
  Science and Engineering, National University of Singapore, 
  117542, Singapore}
\affiliation{Center for Phononics and Thermal Energy Science,
  School of Physics Science and Engineering, Tongji University,
  200092, Shanghai, China}
   
  \author{H. Eugene~Stanley}
  \affiliation{Center for Polymer Studies and Department of
  Physics, Boston University, Boston, MA 02215} 
  
  \author{Shlomo Havlin}
  \affiliation{Department of Physics, Bar-Ilan University, 52900
  Ramat-Gan, Israel}
  \author{Lidia A. Braunstein}
  \affiliation{Center for Polymer Studies and Department of
  Physics, Boston University, Boston, MA 02215}
    \affiliation{Instituto de Investigaciones Fisicas de Mar del Plata (IFIMAR), Universidad
  Nacional de Mar del Plata-CONICET, Funes 3350, (7600) Mar del Plata,
  Argentina}

\begin{abstract}

Although the many forms of modern social media have become major
channels for the dissemination of information, they are becoming
overloaded because of the rapidly-expanding number of information
feeds.  We analyze the expanding user-generated content in Sina Weibo,
the largest micro-blog site in China, and find evidence that popular
messages often follow a mechanism that differs from that found in the
spread of disease, in contrast to common believe. In this mechanism, an individual with more friends
needs more repeated exposures to spread further the
information. Moreover, our data suggest that in contrast to epidemics,
for certain messages the chance of an individual to share the message
is proportional to the fraction of its neighbours who shared it with
him/her.  Thus the greater the number of friends an individual has the
greater the number of repeated contacts needed to spread the
message, which is a result of competition for attention. We model this process using a fractional susceptible infected
recovered (FSIR) model, where the infection probability of a node is proportional to its fraction of infected neighbors. Our findings have dramatic implications for
information contagion. For example, using the FSIR model we find that
real-world social networks have a finite epidemic threshold. This is
in contrast to the zero threshold that conventional wisdom derives
from disease epidemic models. This means that when individuals are
overloaded with excess information feeds, the information either
reaches out the population if it is above the critical epidemic
threshold, or it would never be well received, leading to only a
handful of information contents that can be widely spread throughout
the population.

\end{abstract}

\maketitle
%

\section{Introduction}

Because of the expanding size of such online social networks (OSNs) as
Facebook and Twitter, modern media carry an enormous amount of user
generated content. As their impact on society is increasing, much
interest is now being focused on the spreading mechanism in social
networks.  In order to understand the mechanisms underlying information
diffusion, many studies involve analyzing large amounts of empirical
data
\cite{adar2005,anagnostopoulos2008,gruhl2004,cha2007,guo2008,lerman2010,bakshy2012},
and others formulate predictions of how popular a particular piece of
information will become \cite{galuba2010,lee2010}. The susceptible
infected recovered (SIR) model \cite{newman2002} of disease epidemics is
frequently used to model the spread of information. 

Although disease, opinion, and information spreading all share
significant similarities, fundamental differences remain. In the spread
of disease \cite{bailey1975, Satorras2001, Moreno2002, Meloni2012, Maziar2007}, every person coming in contact with an
infected individual has the same probability of being infected, and the
infected individual continues to infect others until it no longer has
the disease. In contrast, an individual with many friends may not share
a held opinion if only a few of the friends agree with the
opinion. Indeed, in many opinion models \cite{sznajd2000, Valdez2013, liggett1999, lambiotte2008, galam2002, krapivsky2003, nowak1990, deffuant2000, hegselmann2002, bunde1991, shao2009,li2011, watts2002,liu2012}
individuals strongly tend to conform to the majority opinion of their
friends. The spreading of information on OSNs is similar to that in
epidemic and binary-choice opinion models, but the detailed mechanisms
can differ. Studies \cite{holthoefer2013a,holthoefer2013b} have shown that the epidemic models do not reproduces certain empirical statistics observed in information spreading, and modifications on `inactive' and `ignorant' behaviours could significantly improve model results, and the influence of super-spreaders are present to ensure the matching statistics between models and empirical data. Social experiments have found that individuals
often adopt new social behaviors when they are strongly influenced
by repeated signals from friends \cite{centola2010}, and extensive
empirical study of Facebook found that the predominant component of
Internet content spreading is the influence of ``weak'' links, e.g., the
viewing of content generated by individuals with whom the viewer has had
no interaction. This suggests that these weak links play a much more
important role in information diffusion in OSNs than in face-to-face
social networks where there is social interaction
\cite{granovetter1983}. This same study \cite{granovetter1983} and
others \cite{hodas2012} also found that most of the spreading of online
information occurs within the first day of its posting, indicating a
short lifetime and a decaying rate of diffusion similar to that in the
spread of disease.  In addition to the social reinforcement behaviors
discovered in Ref.~\cite{centola2010}, Ref.~\cite{hodas2012} also found
that highly connected nodes (individuals with many friends) are less
likely to spread (pass on) incoming information.

Although most empirical analyses focus on large data sets and use
average statistics as a reference when determining universal spreading
mechanisms, there is evidence \cite{romero2011} that spreading modes
differ as information types differ and that they are also influenced by
the number of linked ``friends'' a user has.  Here we examine a set of
popular messages on the Chinese microblog site Sina Weibo to determine
the number of repeated message ``shares'' an individual needs to receive
before they in turn share the message. We find that much of the viral
spreading of popular messages follows a fractional SIR model, i.e., an
individual with a large number of friends needs more repeated signals
before sharing a message than one with a few number of friends.  The
analysis also suggests that the probability that an individual will
share a message is proportional to the fraction of its friends that have
shared it with him or her. This mechanism leads to a phase transition
behavior in the spreading, i.e., messages with attractiveness below a
critical threshold will not spread to a significant fraction of the
population of the OSN, and those above the critical threshold will.

\begin{figure}[htb]
\includegraphics[width=0.9\linewidth]{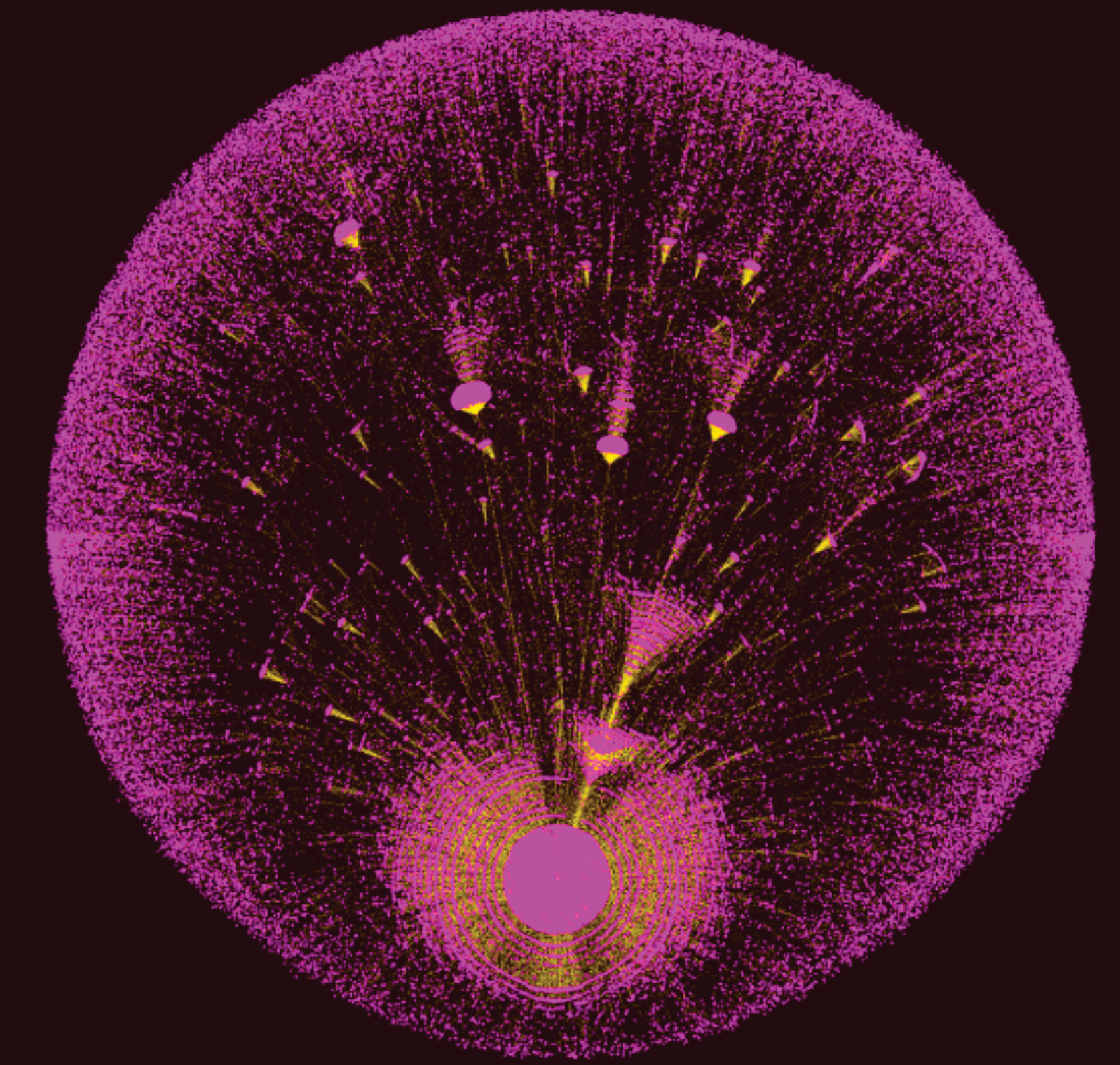}
\caption{Graphical representation of a diffusion tree on Sina
  Weibo. This is a graphic representation for one of the popular
  messages on Weibo, and the branching process can be seen. It has been
  shared more than 190,000 times. The purple dots represent users who
  have shared the message, and green lines represent the paths through
  which the messages are spread. If a user A shared the message from
  user B, a green line is drawn between both. The clusters of purple
  dots corresponds to a user with a large number of followers who have
  shared the message from this user.}
\label{spreadingtree}
\end{figure}

\section{Empirical Motivation}

\subsection{Data Source}

We obtain our data from the micro-blog site Chinese Sina Weibo
(www.weibo.com), one of most popular social media channels in China. It
is similar to Twitter and by the end of 2012 had more than 40,000,000
active users \cite{weibo}. A user can view messages that other users post but cannot
send messages to them unless the other users elect to ``follow'' this
user. The messages are limited to 140 Chinese characters. The number of people a
user with a free account can follow is limited to 2000. (A tiny number
of users elect to have the paid account option that allows a higher
limit.)  Figure~\ref{spreadingtree} shows the transmission pattern for
one of the popular messages on Weibo. The purple dots are users who have
shared the message, and the green lines are the paths through which the
message has spread. The spreading is a branching process that originates
from a first node near the bottom of the figure.

In many empirical studies of information diffusion and user behavior in
social media, aggregate statistics have been derived using a large
amount of information content \cite{bakshy2012,hodas2013}. Although this
approach has the advantage of having large data sets and reliable
statistics, it ignores the heterogeneity of information content and its
effect on any findings.

In order to examine how message transmission dynamics vary as the type
of message varies, we select the most popular 286 messages that were
shared at least 40,000 times in December 2012. Because we
focus solely on these popular messages, we avoid the need to identify
and filter spam \cite{ghosh2011}. For each message, we obtain the
diffusion process by identifying every node (every user account) that
has shared the message and recording the time of each sharing. For each
node in this diffusion tree, we get the number of times it receives the
same message from friends before it in turn shares the message. We do
this by combining the network structure data with the sharing-time data.

\subsection{Empirical Observations}

We first define three quantities:

\begin{itemize}

\item[{(i)}] The number of users $N_i$ sharing message $i$, which in
  epidemic spreading terms corresponds to the total number of
  individuals who have been infected and have recovered.

\item[{(ii)}] The fraction of users sharing the same message as
  $R=N_i/N$, where $N$ is the total number of active Weibo users.

\item[{(iii)}] The total number of followees of user $j$ (the total
  number of users being followed by $j$) who share share a message with
  $j$ before $j$ shares the message is defined as $k_j^-$. In the
  epidemic model this corresponds to the number of infected neighbors of
  node $j$ before $j$ is infected. The total number of users that user
  $j$ follows is defined as $k_j$. For each message, we denote the
  average of $k_j^-$ of all the users sharing this same message as
  $\langle k_- \rangle$.

\end{itemize}

\begin{figure}[htb]
 \includegraphics[width=\linewidth]{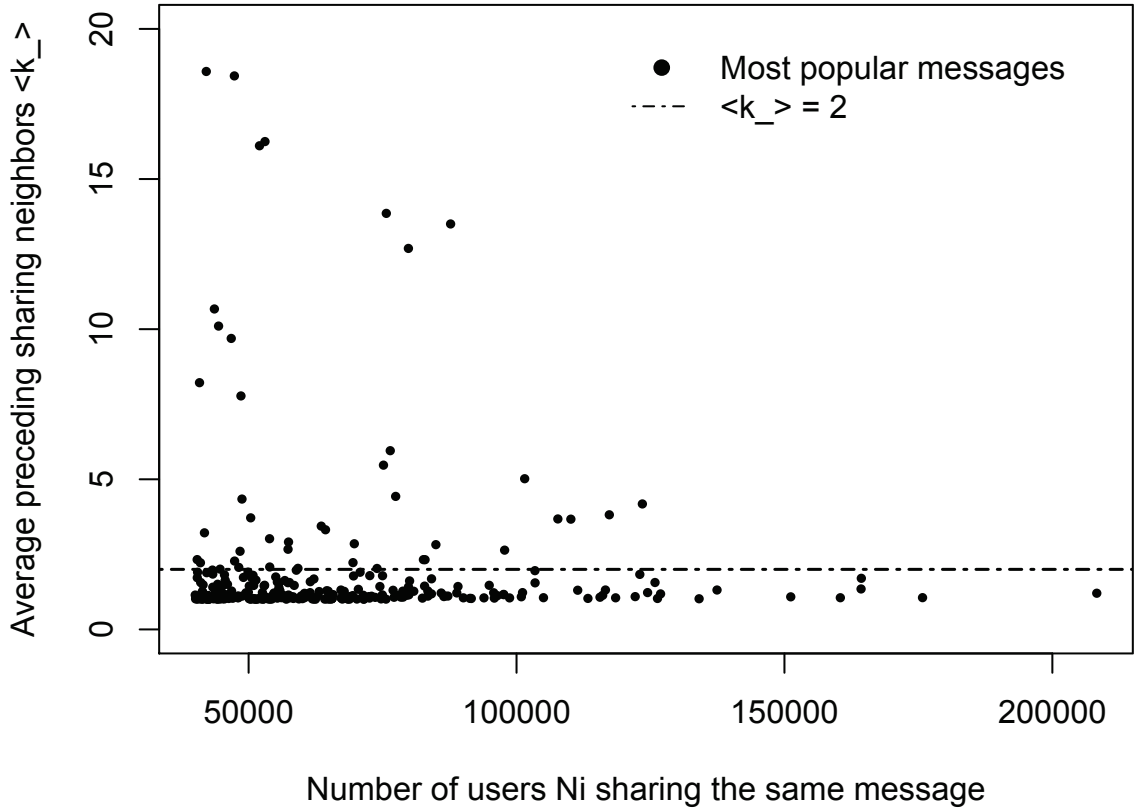}
 \caption{Plot of $\langle k_-\rangle$ for each of the the most popular
   messages as a function of $N_i$. Messages are the most popular ones
   from Sina Weibo in December 2012. Majority of the most popular
   messages have average preceding sharing neighbors smaller than 2,
   which contradicts the SIR model. This hints that the mechanism
   might be different from SIR models for disease spreading for
   scale-free network of OSN type.}
\label{smallbig}
\end{figure}

Figure~\ref{smallbig} shows that in the majority of these messages have
an average value $\langle k_-\rangle<2$. This is in contrast to the SIR
model of scale-free networks in which, as we examine in more detail
below, $\langle k_- \rangle>2$.

Thus the hypothesis of our new model is that the spread of information
in an OSN differs from the spread of a disease during an epidemic. In an
OSN users pay only limited attention to incoming information and as the
number of their contacts increase this attention to each contact decreases further.  In the
following section we test this hypothesis by comparing the outcome
from our model with real-world data.

\section{The Fractional SIR (FSIR) Model}

In our fractional SIR (FSIR) model we assume that, as the number of
friends a user has increases, the number of enforcements from these
friends of a particular message required before the user passes the
message on (``spreads the infection further'') also increases.  Because
the total amount of attention a user can pay to the OSN is limited, we
hypothesize that the amount of influence from each neighbor is inversely
proportional to the total number of contacts (followees) the user has.

Evidence exists \cite{hodas2013friend} that despite the large amount of
information a node with high connectivity $k$ can {\it receive\/}, the amount of
information it {\it shares\/} is not significantly greater than the
amount of information shared by a node with low connectivity $k$.  Thus
the total amount of attention a node can pay to the total information
received from all of its neighbors is limited, irrespective of the
node's connectivity (there is a cognition limit). If all nodes thus have
approximately the same cognition limit, an increasing overload of
information will cause the higher-degree nodes to pay a decreasing
amount of attention to information received from each of its contacts.

Our model assumes a network of $N$ nodes, each representing a user
account on the OSN in which the number of contacts $k$ of a node is
given by a degree distribution $P(k)$. In the real-world Weibo network
the links are directed because the follower of a user account can see
its posts but the user account cannot see the posts of its followers. For simplicity we
use undirected networks in our analysis. Reference~\cite{rodriguez2014}
shows that on the microblog site Twitter the users sharing the most
messages are also the ones receiving the most messages despite the fact
that the number of their followers differ from their number of
followees.  Hence an undirected network is a good approximation of
message flow in a social network.  The spreading mechanism in our model
is as follows:

\begin{enumerate}

\item A node has information (is infected) at step $t=0$. At a
  subsequent time $t$, for each node $i$ that has not shared the
  information (has not been infected but is susceptible) but has friends
  that have (have been infected), the probability that $i$ will
  subsequently share the information (the infection) from each infected (sharing) friend (followee) is $\gamma/k_i$, where $k_i$ is the degree of node $i$.

\item $\tau$ time steps after sharing (infection), an infected individual recovers and cannot be infected again (is no longer visible to its neighbors on the
  Internet).  Hence the probability a susceptible node will be infected
  by a sharing friend is $T_{k_j}=1-(1-\gamma/k_j)^\tau$, where $k_j$ is the
  degree of the susceptible node $j$.

\item At the final steady state nodes can no longer be infected, and all
  of the nodes in are in the recovery or susceptible state.  The message
  is no longer being shared.

\end{enumerate}

Here $\gamma$ is the intrinsic attractiveness of the information, and
$\tau$ is the visible duration on the feeds of a user's
followers.

Step one of our model differs from the corresponding step in the SIR
model. In the SIR model, the probability of infection is $\gamma$ (and
is independent of the degree of the susceptible node), and the effective
probability of infection is $T=1-(1-\gamma)^\tau$ and is independent of
$k$. In contrast, the infection probability in our FSIR model is
determined by the individual node degree $\gamma/k$. Thus a node of
degree $k$ can be infected with probability
$T_k=1-(1-\gamma/k)^\tau$. In a real-world OSN, $k \gg \gamma$, and thus
$T_k \approx \gamma \tau /k= \Gamma/k$, with $\Gamma\equiv \gamma
\tau$. In the case of SIR, $T \approx \Gamma $ for any node of any
degree $k$.

We perform FSIR and SIR simulations on a scale-free network of size
$N=100,000$ and degree distribution $P(k) \sim k^{-\lambda}$, with
$\lambda=2.5$, which is the approximate empirical degree distribution of
real-world OSNs. We fix the average degree at $\langle k \rangle=50$ to
stay close to real-world OSNs. Since the popularity of the empirical
data is typically less than 0.5\% of the total population of users (200,000 shares out of 40,000,000 active users),
i.e., $R \approx 0.5\%$, we use $\Gamma$ values in our simulations that
give similar $R$ values, which corresponds to $N_i \approx 500$. We run
5000 realizations for both the FSIR and the SIR model, and select out
the realizations in which $N_i>200$, which corresponds to the $R$ values
of the highly popular messages selected from real-world data.

\begin{figure*}[htb]
\includegraphics[width=0.9\textwidth]{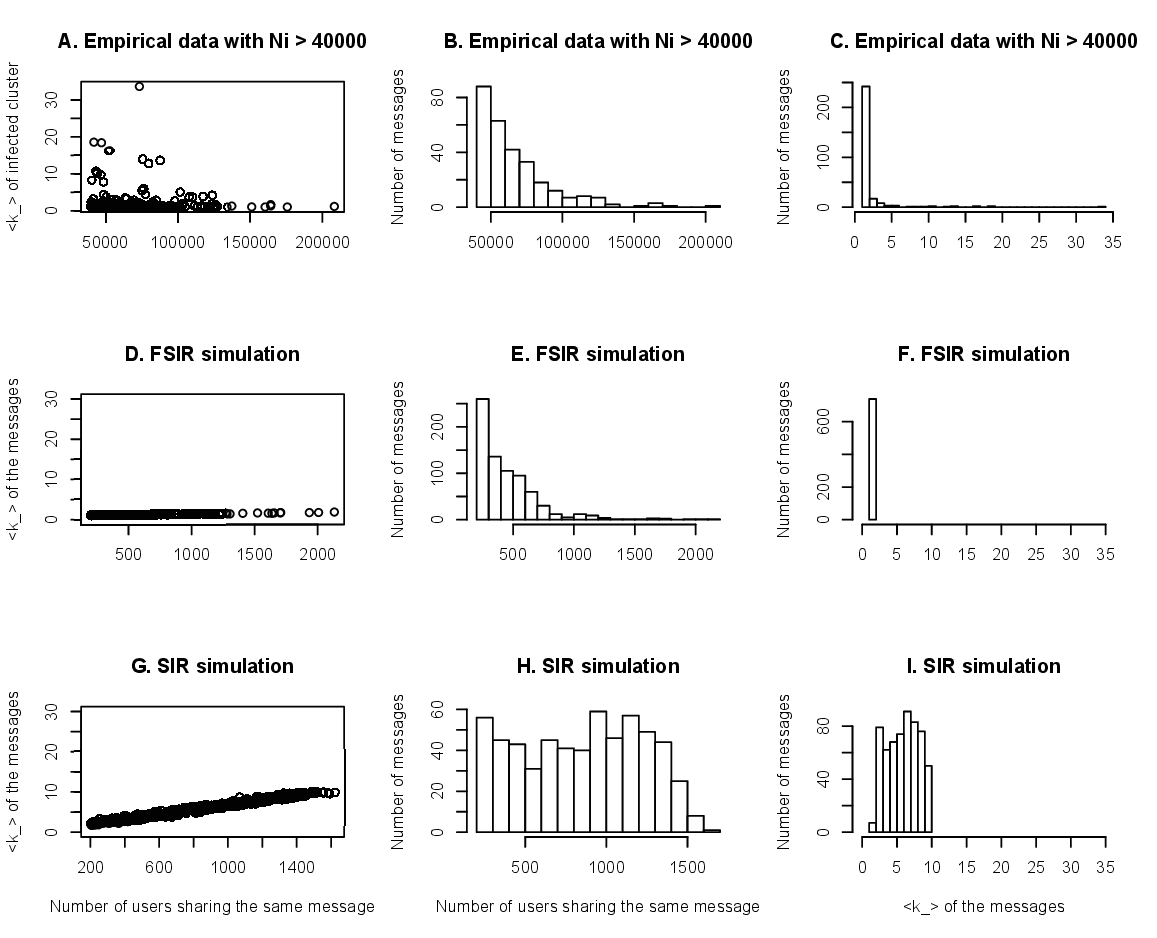}
\caption{Comparison between empirical data and simulation results of
  FSIR and SIR models. In the simulation, the total number of nodes is
  $N=100,000$. The degree distribution is chosen at $P(k) \sim
  k^{-2.5}$, $k_{min}= 20$ and $\langle k \rangle \approx 50$. With
  these parameters this distribution is close to the empirical
  distribution of OSNs. The chosen parameter $\Gamma=0.7$ for FSIR and
  $0.15$ for SIR is such that the fraction of infected nodes $R=N_i/N
  \approx 0.5\%$, is close to the empirical $R$ values of the most
  popular messages. As we can see, for empirical and FSIR, $\langle k_-
  \rangle$ values does not change significantly with $N_i$ as seen in A
  and D. However, as seen in G, SIR shows a clear increase of $\langle
  k_-\rangle$ as a function of $N_i$. As seen in D and F, most of the
  real messages and FSIR result have $\langle k_- \rangle <2$, meaning
  they are small outbreaks below epidemic threshold. But in I, we see
  that for SIR simulations, $\langle k_- \rangle>2$. This means the
  messages are epidemics spreading rather than small outbreaks. In B and
  E, we show that both real data and FSIR have similar distributions of
  $N_i$ values, with number of messages decreases significantly with
  increasing popularity $N_i$. This is in contrast to SIR result in
  H. Hence through simulations, we show that FSIR captures the
  statistics of real data from popular messages, yet SIR does not.}
\label{comparison}
\end{figure*}

Figure~\ref{comparison} shows the simulation results for FSIR and SIR
models with $\Gamma$ values that give outbreak fractions around
$0.5\%$. In order to match the empirical $R$ value of $0.005$, we assume
$\Gamma=0.7$ for FSIR and $\Gamma=0.15$ for SIR. Comparing A, D, and G
in Fig.~\ref{comparison}, we see that FSIR produces $\langle k_-
\rangle$ values below $2$ that do not change significantly with $N_i$,
which is similar to the empirical observation in A.  In contrast, the
range of $\langle k_- \rangle$ values in the SIR model is mostly above 2
and the values increase with $N_i$. In Sec.~V we show that the FSIR has
a phase transition at $\Gamma_c\approx 1$. This suggests that the most
popular messages will spread below a critical threshold, which is
similar to the FSIR mechanism but not to the SIR.  Comparing B, E, and
H, we see that the FSIR model produces a distribution of $N_i$ that
decreases as $N_i$ increases, which similar to the empirical
distributions. In contrast, the distribution of the SIR model is
uniform.  Figures~C, F, and I show that the number of messages with
values of $\langle k_- \rangle<2$ in the empirical data clearly resemble
those of the FSIR simulations, not the SIR simulations in which most of
the $k_-$ values are above 2. Thus the FSIR simulations capture the
statistical properties of real data closely and the SIR simulations do
not.

Our analysis thus strongly suggests that highly popular messages spread
with a mechanism closely modeled by FSIR simulations but not
conventional SIR simulations.  This supports the hypothesis that a user
with a larger number of neighbors will be proportionally less influenced
by each of them, a clear contrast to SIR behavior.  Note however that
there are a small number of messages in which $\langle k_- \rangle >
2$. Perhaps this indicates that while most of the highly popular
messages will follow the FSIR mechanism below the critical threshold,
there are some that are above it. These do not tend to be among the most
widely spread, however, and this could be due to their overall short
lifespan.  For example, a message about an upcoming election will be of
interest to users prior to the election day. After the election the
message disappears.

\section{The phase transition of the FSIR model}

\begin{figure}[thb]
\includegraphics[width=0.9\linewidth]{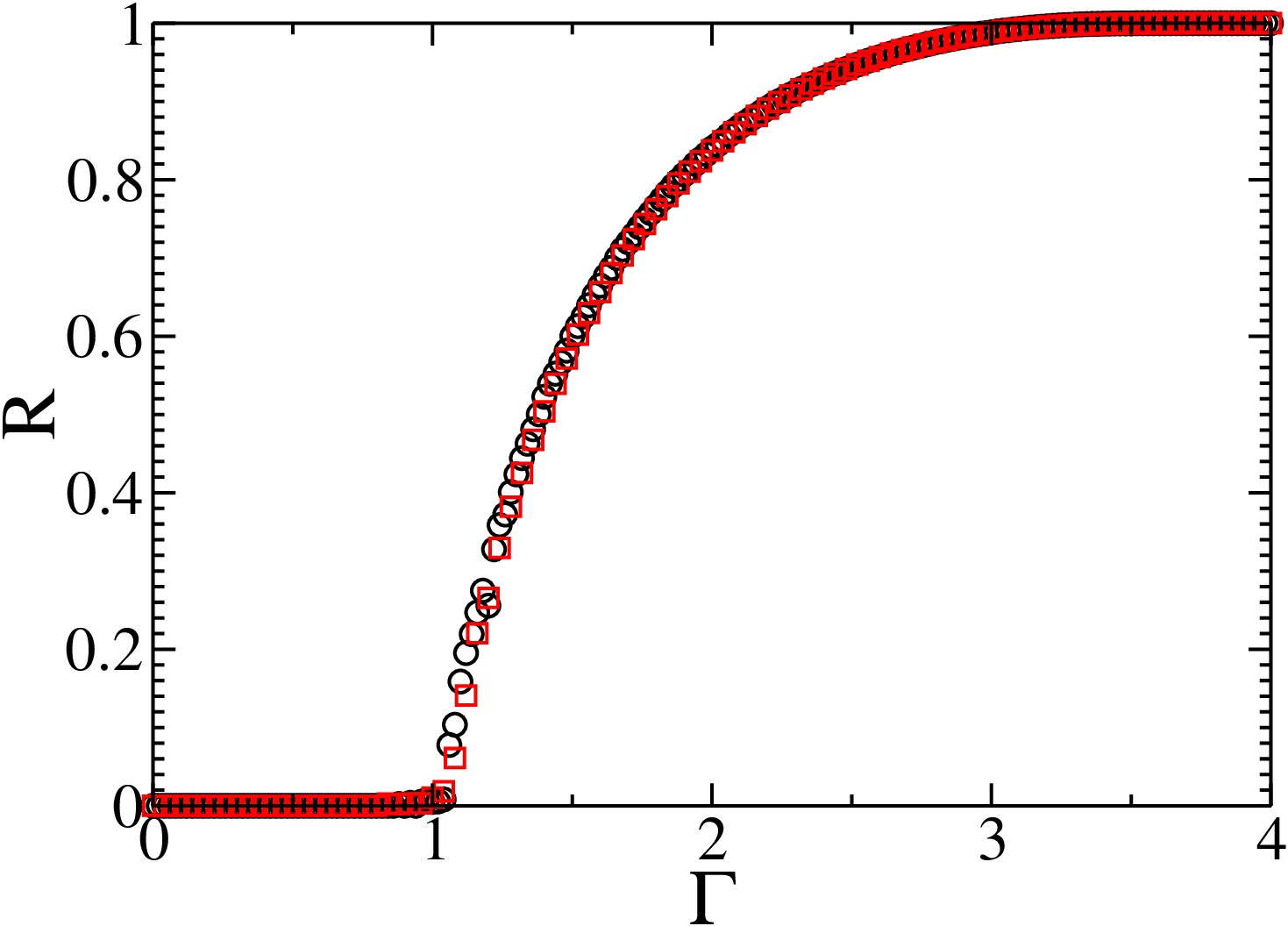}
\caption{$R$ as a function of $\Gamma$ for $\tau=2$ with $N=10^5$ on SF
  networks with $\lambda=2.5$ ($\bigcirc$) and $\lambda=3$ ($\Box$) and
  $k_{min}=20$. As shown in the plots FSIR has a critical point around
  $\Gamma_c \approx 1$, independent of the degree distribution. The same
  critical value $\Gamma_c$ is observed for other degree distributions
  with different parameter settings.}
\label{NiAll}
\end{figure}

We next analyze the impact of the FSIR mechanism on information
epidemics.  On scale-free networks (or networks with broad degree
distributions) the SIR mechanism has a critical threshold for epidemics
of $\Gamma_c \approx 0$. This means that any disease is able to infect a
considerable proportion of a population regardless of its intrinsic
ability to spread.  As suggested above, the spread of information in a
OSN follows the FSIR mechanism, and thus, as we will show below, the
critical threshold is $\Gamma_c \approx 1$.  Below this critical
$\Gamma_c$ threshold, information ``outbreaks'' can develop but not
information ``epidemics''. Above this threshold the information can
diffuse to a much wider finite fraction of the user
population. Figure~\ref{NiAll} plots the FSIR simulation results for
scale-free networks with two different values of $\lambda$. In these
simulations $N=100,000$, $\tau=2$, and $k_{\rm min} =20$ for
$\lambda=2.5$ and $\lambda=3$.  As $\Gamma$ increases the FSIR shows a
critical phase transition at $\Gamma_c \approx 1$, which differs from
the SIR \cite{Vespignani2001} in which $\Gamma_c \approx 0$ for $N \to
\infty$.

For a random network with degree distribution $P(k)$ in the FSIR model,
$T_k \approx \Gamma/k$.  Using the generating function formalism
\cite{newman2002, braunstein2007, callaway2000, kenah2007} for an
inhomogeneous $T_k$, we can derive the probability $f_\infty$ that a
branch of infection reaches infinity using the self-consistent equation
\begin{equation}\label{Eq.f}
f_\infty=1-G_1(1- \frac{\Gamma}{k} f_\infty),
\end{equation} 
where 
\begin{equation}
  G_1(1-\frac{\Gamma}{k} x) =\sum_k \frac{k P(k)}{\langle k \rangle} (1-
  \frac{\Gamma}{k} x)^{k-1}.
\end{equation} 

\begin{figure}[htb]
\includegraphics[width=0.9\linewidth]{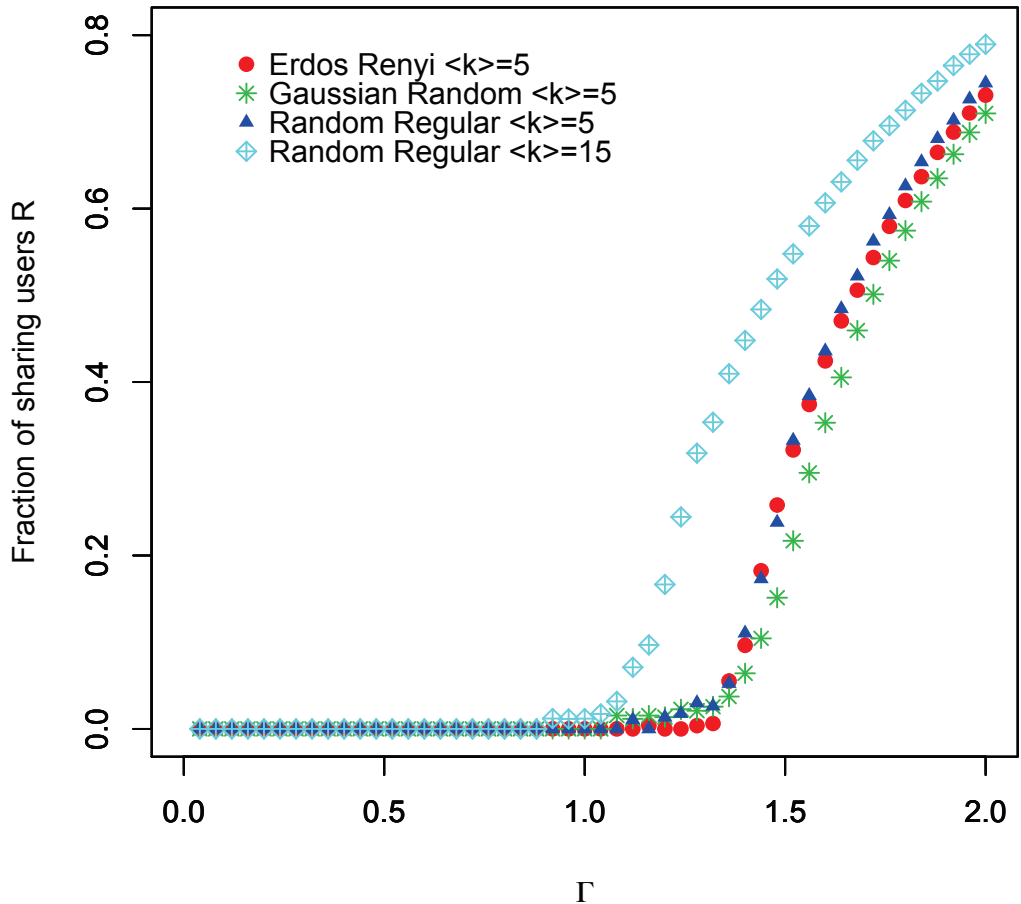}
\caption{Plot of $R$ as a function of $\Gamma$ for different types of
  random networks. Regardless of the types of networks, the critical
  threshold for phase transition $\Gamma_c$ is always larger than
  $1.0$. For networks with larger $\langle k \rangle$ values, $\Gamma_c$
  is closer to $1.0$ as predicted by Eq.~(\ref{GammaC}).}
\label{AllR}
\end{figure}

The non-trivial solution of Eq.~(\ref{Eq.f}) gives the value of $f_\infty$
for a given value of $\Gamma$ that corresponds to the intersection
between the identity and the right side of Eq.~(\ref{Eq.f}).  At
criticality there is only one root that corresponds to
$f_\infty=0$. Thus the left side of Eq.~(\ref{Eq.f}) must be tangent to
the identity at $f_\infty=0$.  In other words, the derivative of
Eq.~(\ref{Eq.f}) evaluated on this root must be one.  Thus
\begin{equation}
 \Gamma_c \;  \frac{\sum_k (k-1) P(k)} {\langle k \rangle}=1,
\label{TheoryCritical}
\end{equation}
and
\begin{equation}
\Gamma_c = \frac{\langle k \rangle}{\langle k \rangle -1},
\label{GammaC}
\end{equation}
independent of the degree distribution $P(k)$ (see Fig.~\ref{AllR}).  In
a real OSN in which $\langle k \rangle$ is of the order of $100$,
$\Gamma_c \approx 1$.

The fact that $\Gamma_c =1$ has a significant impact on information
epidemics means that, in an OSN, most messages cause small information
outbreaks at best, and only highly attractive and interesting messages
($\Gamma>1$) reach a significant fraction $R$ of the user population.
The $\Gamma$ value of individual messages depends not only on how
attractive, interesting, or novel a particular message is, but also on
how attractive, interesting, or novel all the other messages being
shared during the same period are. The novelty and attractiveness of all
messages can theoretically double, i.e., the $\Gamma$ value of every
message can double, but the attention of the users cannot.  Thus what
matters is the $\Gamma$ value of a particular message as compared to the
$\Gamma$ values of all the other messages, because all messages are
competing for attention from the same user population.  In other words,
$\gamma$ values can indicate the weighted novelty of a message compared
to all other messages. If there are too many messages with similar a
level of novelty, it is probable that none of them will have $\Gamma>1$,
and thus none will prevail. When there is an individual message that
much more interesting than all the other messages, it may capture a considerable fraction of the user population.

\section{Conclusion and Discusion}

Note that the FSIR mechanism is more likely to be present in situations
of information overload. When there is an information overload, users
select what they will share and what they will ignore. This is what
makes information spreading distinctively different from the spread of
disease. A piece of information competes with all other pieces of
information for the attention of users (``nodes''), but disease
epidemics can utilize every opportunity to spread to other individuals
irrespective of the presence of other diseases. In fact, an individual
who has been exposed to many diseases is {\it more likely\/} to be
infected. In contrast, an individual inundated by an overload of
messages is {\it less likely\/} to view, remember, or pass on any of
them.

Information overload also shortens the visibility duration for popular
messages.  Because messages come to a user every day, new messages
appearing above old messages, an information overload means any message,
however popular, will rapidly lose its visibility, thus effectively
shortening the $\tau$ value. Even extremely popular units of information
content die quickly. An outstanding example was the immensely popular
but short-lived ``Gangnam Style'' music video that quickly spread across
the world---receiving over one billion views on Youtube---and then soon
after lost its popularity. There are many other current examples in pop
culture and on the Internet of subjects that quickly become world-wide
topics and then quickly disappear.  In marketing also, many products
compete with each other and overload consumers with advertising
messages, which may influence them, and their choices may in turn
influence their friends, but the result is usually a handful of brands
dominating a market with the rest of the brands fighting for survival.

Because this critical phenomenon is present irrespective of network
structure, and because people encounter information overload in many
settings, the FSIR mechanism and its associated phase transition
phenomenon are present in many other real-world contexts.  As
information spreads across an OSN according to either the SIR or FSIR
mechanism, any attempt to predict the popularity of a unit of
information content must first determine which mechanism is present.
Only then can business or government institutions, for example,
obtain useful insights into the behavior of information speading. 

The bigger question raised from this work is what determines the
spreading mechanism of a message, and what contributes to its novelty
value ($\Gamma$ value). Related empirical studies \cite{aral2011, aral2012} on social contagion of products provided valuable insights into some important features of marketing campaigns that induces viral spreading. Our work provides a systematic framework to mathematically relate such features to adoption rates.
We believe that, combining our mathematical framework with large pool of data and the right methodology from machine learning and text mining, more can be achieved in this field with immense social values.

\section{Acknowledgements}

This work is supported by the Singapore NSF grant ``Econophysics and
Complex Networks'' (R-144-000-313-133).  We also wish to thank ONR,
DTRA, NSF the European MULTIPLEX, LINC and CONGAS projects and the
Israel Science Foundation for financial support. LAB thanks to UNMdP
and FONCyT, Pict 0429 for financial support.

\bibliographystyle{unsrt}

\end{document}